\documentclass[prl,twocolumn,showpacs,amsmath,amssymb]{revtex4}

\usepackage{wasysym}
\usepackage{graphicx}
\usepackage{dcolumn}
\usepackage{bm}

\begin{document}

\title{Long-range correlations of density in a Bose-Einstein condensate expanding in a random potential\\}

\author{N. Cherroret}
\author{S.E. Skipetrov}
\affiliation{
Universit\'{e} Joseph Fourier, Laboratoire de Physique et Mod\'{e}lisation des Milieux Condens\'{e}s, CNRS, 25 rue des Martyrs, BP 166, 38042 Grenoble, France
}

\date{\today}

\begin{abstract}
We study correlations of atomic density in a weakly interacting Bose-Einstein condensate, expanding diffusively in a random potential. We show that these correlations are long-range and that they are strongly enhanced at long times. Density at distant points exhibits negative correlations. 
\end{abstract}


\maketitle

The behavior of Bose-Einstein Condensates (BECs) in disordered potentials has attracted growing interest of physicists during the last few years. In particular, expansion of BECs in random potentials has been investigated in detail \cite{experimental_papers,Sanchez-Palencia,Shapiro,Sergey,Billy,Roati}. The main interest of using BEC systems to study disorder-related phenomena is that their physical parameters (such as the number of atoms, the parameters of the random potential, or the strength of inter-atom interactions) can be controlled quite precisely.
During the last two years, considerable efforts were undertaken to study the atomic density $n(\textbf{r},t)$ averaged over disorder in 1D \cite{Sanchez-Palencia}, 2D \cite{Shapiro}, and 3D \cite{Shapiro,Sergey} geometries, with special interest in the phenomenon of Anderson localization \cite{Sanchez-Palencia,Sergey,Billy,Roati}.
At the same time, little is known about statistical fluctuations of $n(\textbf{r},t)$ due the randomness of the potential. Recently, Henseler and Shapiro \cite{Henseler} have shown that a BEC expanding in a random potential is characterized by a complicated, highly irregular density pattern reminiscent to what we know as ``speckle'' in optics \cite{goodman}.  According to Ref.\ \cite{Henseler}, multiple scattering from the potential completely randomizes $n(\textbf{r},t)$ and reduces the correlation length of atomic speckle pattern to a value of the order of the healing length of the initial trapped condensate, which is the minimal possible correlation length for a coherent matter wave.
 The purpose of this Letter is to show that the macroscopic coherence of the condensate gives rise to genuine interference effects that were ignored in Ref.\ \cite{Henseler}. These interference effects are similar in nature to those leading to Anderson localization of the condensate at large disorder \cite{Sergey}. They result in stronger fluctuations and long-range correlations of $n(\textbf{r},t)$, akin to the long-range correlations of conductance fluctuations in disordered metals \cite{kane88} and the long-range correlations of intensity in optical speckle patterns \cite{feng91,Berkovitz}. For a BEC expanding inside an optical waveguide, the long-range correlations grow in absolute value with time and become dominant in the long-time limit. They can take both positive (for relatively close points) and negative (for distant points) values.

\begin{figure}[t]
\includegraphics[width=6cm]{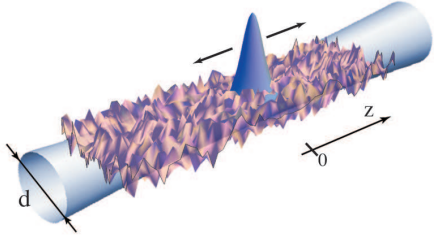}
\caption{\label{BEC_expanding} 
(color online). Cartoon of a BEC expanding in a 3D random potential and confined transversally to a waveguide of typical diameter $d$. We assume $2 \pi/k_\mu \ll d \lesssim \ell$. 
}  
\end{figure}

We consider a dilute BEC of $N \gg 1$ atoms of mass $m$ inside an infinitely long optical waveguide of diameter $d$ and cross-section $A = \pi d^2/4$, parallel to the $z$ axis and described by a 2D potential $V_{\perp}(x,y)$, see Fig.\ \ref{BEC_expanding}. 
The waveguide geometry is rather popular in BEC experiments and was, in particular, used in the recent work on Anderson localization \cite{Billy}.
In the longitudinal direction, the condensate was initially confined by a tight 1D parabolic trap potential $V_z(z) = m \omega_z^2 z^2/2$ that has been switched off to let the condensate expand along the $z$ axis.
  A time $T \gtrsim 1/\omega_z$ later the role of interactions between atoms in the condensate becomes negligible \cite{Pitaevskii,Sanchez-Palencia,Billy} and a weak 3D random potential $V(\textbf{r})$ is switched on; we refer to this moment as $t = 0$. $V(\textbf{r})$ is assumed to have a white-noise Gaussian statistics: $\overline{V(\textbf{r})}=0$ and $\overline{V(\textbf{r})V(\textbf{r}^{\prime})}=u\delta(\textbf{r}-\textbf{r}^{\prime})$, where the horizontal bar denotes averaging over realizations of the random potential. The associated mean free path is $\ell=\hbar^4\pi/um^2$ \cite{Akkermans} and the ``weakness'' of the random potential is quantified by a condition $k_ {\mu} \ell \gg 1$, where $k_{\mu} = \sqrt{2 m \mu}/\hbar$ and $\mu$ is the chemical potential of the initial trapped condensate. 
For $t > 0$, the macroscopic wave function of the condensate $\psi(\textbf{r}, t)$ obeys the  linear Schr\"{o}dinger equation \cite{Pitaevskii}:
\begin{equation} \label{Pitaevskii}
i \hbar \dfrac{\partial\psi}{\partial t}=\left[-\dfrac{\hbar^2}{2m}\Delta+
V_\perp(x,y)+V(\textbf{r}) \right]\psi.
\end{equation}

In this Letter we assume that the diameter $d$ of the waveguide to which the expansion of the BEC is constrained obeys $2 \pi/k_\mu \ll d \lesssim \ell$. This corresponds to what is known as a ``quasi-1D'' geometry in the multiple-scattering literature \cite{Akkermans}: at distances larger than $\ell$, the average atomic density ${\bar n}(\textbf{r}, t) = \overline{\left| \psi(\textbf{r}, t) \right|^2}$ is described by a 1D diffusion process, unlike the recent work \cite{Sanchez-Palencia,Billy} where 1D (and not 3D) disorder was considered and diffusive propagation didn't appear. 
In the quasi-1D geometry, the localization length at energies of the order of $\mu$ is typically $\xi_\mu \sim \ell (k_\mu d)^2$, exceeding the mean free path $\ell$ by a factor $(k_{\mu} d)^2 \gg 1$. Hence, the condensate expands by diffusion until $\left| z \right| \sim \xi_\mu \gg \ell$ before it starts to be affected by Anderson localization effects. This regime was not accessible in the recent experiment \cite{Billy} and has not been studied theoretically yet.

For weak disorder $k_{\mu} \ell \gg 1$ and at large distances $z \gg \ell$ and long times $t \gg \ell/v_{\mu}$ (with $v_{\mu} = \hbar k_{\mu}/m$ being the typical velocity of an atom with kinetic energy $\mu$), the average atomic density is independent of $x$, $y$ and can be written as \cite{Shapiro,Sergey,Henseler}
\begin{equation} \label{atomic_density_eq}
\bar{n}(z,t)=\int_{-\infty}^{\infty}\dfrac{dk}{2\pi}|\phi(k)|^2P_{\epsilon_k}(z,t),
\end{equation}
where $\left| \phi(k) \right|^2 \propto (1-k^2/2 k_\mu^2) H(1 - \left| k \right|/\sqrt{2} k_{\mu})$ \cite{Pitaevskii,Sanchez-Palencia,Kagan}, with $H(x)$ the Heaviside step function, describes the momentum distribution of atoms in the condensate at $t = 0$, $\epsilon_k = \hbar^2 k^2/2 m$, and $P_\epsilon(z,t)$ is the probability to find a particle of energy $\epsilon$, initially located at the origin, in the vicinity of $\textbf{r} = (x,y,z)$ after a time $t$ \cite{Akkermans}. The Fourier transform of the latter is $P_\epsilon(z, \Omega)=\overline{G_{\epsilon+\hbar \Omega/2}^{}(\textbf{r}, 0) G_{\epsilon-\hbar \Omega/2}^{*}(\textbf{r}, 0)}/2 \pi \nu_{\epsilon}$, where $G_{\epsilon}(\textbf{r}, \textbf{r}^{\prime})$ is the Fourier transform of the Green's function of Eq.\ (\ref{Pitaevskii}) and $\nu_{\epsilon}$ is the local density of states at the energy $\epsilon$. In the hydrodynamic limit $\hbar \Omega\ll\epsilon$, $P_\epsilon$ is a solution of a 1D diffusion equation: $P_\epsilon(z,\Omega)=\exp(-\left| z \right| \sqrt{-i\Omega/D_\epsilon})/2A\sqrt{-i\Omega D_\epsilon}$ \cite{Akkermans}, where $D_\epsilon=v_{\epsilon} \ell/3$ is the diffusion coefficient for atoms at energy $\epsilon$. This yields
\begin{equation}\label{atomic_density_explicit}
\bar{n}(z,t)=\dfrac{N}{A \sqrt{D_\mu t}}f\left(\dfrac{z}{\sqrt{D_\mu t}}\right),
\end{equation}
where $f(x)$ can be expressed through special functions and $f(x) \simeq 0.6$ for $x \gg 1$. 

\begin{figure}[t]
\includegraphics[width=6.2cm]{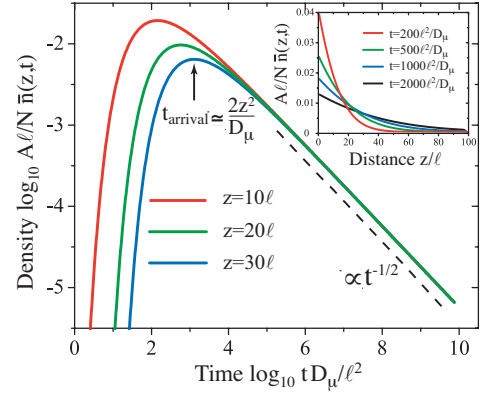}
\caption{\label{atomic_density}
(color online). Ensemble-averaged atomic density $\bar{n}$ of a BEC expanding in a 3D random potential inside a quasi-1D optical waveguide. The main plot shows ${\bar n}$ as a function of time for three different distances $z$. The dashed line is a $1/\sqrt{t}$ asymptote. The inset shows ${\bar n}$ as a function of $z$.
}  
\end{figure}

As the condensate expands, its typical size grows with time according to
$\langle z^2 \rangle \simeq D_{\mu} t$; profiles of atomic density are plotted as functions of $z$ in the inset of Fig.\ \ref{atomic_density}. In the main plot of Fig.\ \ref{atomic_density} we show $\bar{n}$ as a function of time. The density reaches a maximum at the ``arrival time'' $t_{\mathrm{arrival}} \simeq 2 z^2/D_\mu$ and decays as $1/\sqrt{t}$ at long times $t > t_{\mathrm{arrival}}$. This long-time limit is the most interesting regime to which we restrict our analysis from here on. 
It is worthwhile to note that Eq. (\ref{atomic_density_explicit}) breaks down at very long times, when Anderson localization comes into play. Indeed, Eq.\ (\ref{atomic_density_explicit}) predicts ${\bar n} \rightarrow 0$ for $t \rightarrow \infty$, whereas localization will ``freeze'' ${\bar n}$ at values of the order of $N/A \xi_{\mu}$ (for $z < \xi_{\mu}$), starting from some localization time $t_{\mathrm{loc}}$. The latter can be estimated by comparing $N/A \xi_{\mu}$ with the long-time limit of Eq.\ (\ref{atomic_density_explicit}) or, equivalently, by equating $\langle z^2 \rangle$ and $\xi_{\mu}^2$. One obtains $t_{\mathrm{loc}} \sim (\ell/v_{\mu}) (k_{\mu} d)^4$, which shows that a considerable interval of validity exists for Eq.\ (\ref{atomic_density_explicit}) between the typical mean-free time $\ell/v_{\mu}$ and $t_{\mathrm{loc}} \gg \ell/v_{\mu}$, when $k_{\mu} d \gg 1$. 

Let us now study correlations of atomic density in the expanding condensate, which is the primary subject of this Letter. We define the correlation function of density fluctuations  $\delta n(\textbf{r},t)=n(\textbf{r},t)-{\bar n}(\textbf{r},t)$ as
\begin{equation} \label{density-density}
C(\textbf{r}, t; \textbf{r}^{\prime}, t^{\prime}) = 
\dfrac{\overline{\delta n(\textbf{r},t)\delta n(\textbf{r}^{\prime},t^{\prime})}}{\overline{n(\textbf{r},t)} \times \overline{n(\textbf{r}^{\prime},t^{\prime})}}.
\end{equation}
Using the relation
$\psi(\textbf{r},t)=\int d^3\textbf{r}_1 G(\textbf{r}, \textbf{r}_1, t)\phi(\textbf{r}_1)$, we can write the numerator of Eq.\ (\ref{density-density}) as
\begin{eqnarray}
\label{density_density}
&&\overline{\delta n(\textbf{r}, t) \delta n(\textbf{r}^{\prime}, t^{\prime})} =
\frac{1}{(2 \pi \hbar)^4}\times \nonumber \\ 
&&\int
\prod\limits_{j = 1}^4 d^3\textbf{r}_j \;
d\epsilon_j \;
e^{-\frac{i}{\hbar}(\epsilon_1 - \epsilon_2)t - \frac{i}{\hbar}(\epsilon_3 - \epsilon_4)t^{\prime}}\times
\nonumber \\ 
&& K\left(\textbf{r}, t; \textbf{r}^{\prime}, t^{\prime};  \left\{ \textbf{r}_j \right\}, \left\{\epsilon_j \right\} \right) \phi(\textbf{r}_1)\phi^*(\textbf{r}_2)\phi(\textbf{r}_3)\phi^*(\textbf{r}_4),
\end{eqnarray}
where the 6-point kernel $K$ is given by the connected part of a product of 4 Green's functions, averaged over disorder:
\begin{eqnarray}
K &=& \overline{G_{\epsilon_1}^{}(\textbf{r}, \textbf{r}_1) G_{\epsilon_2}^*(\textbf{r}, \textbf{r}_2) G_{\epsilon_3}^{}(\textbf{r}^{\prime}, \textbf{r}_3) G_{\epsilon_4}^*(\textbf{r}^{\prime}, \textbf{r}_4)}
\nonumber \\
&-&  \overline{G_{\epsilon_1}^{}(\textbf{r}, \textbf{r}_1) G_{\epsilon_2}^*(\textbf{r}, \textbf{r}_2)} \times \overline{G_{\epsilon_3}^{}(\textbf{r}^{\prime}, \textbf{r}_3) G_{\epsilon_4}^*(\textbf{r}^{\prime}, \textbf{r}_4)}.
\label{kernel}
\end{eqnarray}

\begin{figure}[t]
\includegraphics[width=4cm]{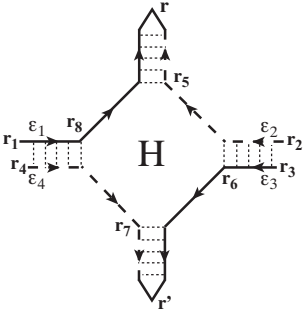}
\caption{\label{Diagram_C2} Diagram for the kernel $K_2$ generating the long-range correlation function of density fluctuations $C_2$ through Eq.\ (\ref{density_density}). The solid and dashed lines represent the Green's functions $G$ and it complex conjugates $G^*$, respectively. The two parallel $G$-lines connected by dotted ``ladders'' symbolize averages of products of two Green's functions, $\overline{GG^*}$. $H = (\ell^5 m^3 /3 \pi \hbar^3 \epsilon_1) \boldsymbol{\nabla}_{\textbf{r}_5} \cdot \boldsymbol{\nabla}_{\textbf{r}_7}$ is the Hikami box \cite{Gorkov,Hikami}. Integration should be performed over $\textbf{r}_5 = \textbf{r}_6 = \textbf{r}_7 = \textbf{r}_8$.}  
\end{figure}

The largest contribution to $K$ is obtained by decoupling $\overline{GG^*GG^*}$ in the first line of Eq.\ (\ref{kernel}) as if $G$ were a circular complex Gaussian random field:
$K_1 = \overline{G_{\epsilon_1}^{}(\textbf{r}, \textbf{r}_1) G_{\epsilon_4}^*(\textbf{r}^{\prime}, \textbf{r}_4)} \times
\overline{G_{\epsilon_3}^{}(\textbf{r}^{\prime}, \textbf{r}_3) G_{\epsilon_2}^*(\textbf{r}, \textbf{r}_2)}$. When $K = K_1$ is inserted into Eqs.\ (\ref{density_density}) and (\ref{density-density}), the short-range correlation function $C_1$ studied in Ref.\ \cite{Henseler} is obtained. $C_1$ is of order 1 for $\textbf{r} = \textbf{r}^{\prime}$, but rapidly decays to zero already for $\left| \textbf{r} - \textbf{r}^{\prime} \right| \sim 1/k_{\mu}$.  
The long-range part of the correlation function --- $C_2$ --- can be obtained by using a next-order contribution to $K$: $K_2$ given by the diagram depicted in Fig.\ \ref{Diagram_C2}.
This diagram represents an interference process between four matter waves that propagate in pairs to some point, where they interchange partners before continuing to the measurement points $\textbf{r}$ and $\textbf{r}^{\prime}$.
A proper treatment of such a crossing of wave paths is guaranteed by the use of the so-called Hikami box diagram \cite{Gorkov,Hikami} --- the $H$ box in Fig.\ \ref{Diagram_C2} --- that ensures conservation of particle number in the final result. 
A simplified version of the diagram of Fig.\ \ref{Diagram_C2}, corresponding to equal energies $\epsilon_j = \epsilon_0$ and identical initial points $\textbf{r}_j = \textbf{r}_0$, yields the long-range correlation function of intensity fluctuations for a scalar wave emitted by a point source in a disordered medium \cite{Berkovitz}. Evaluating the diagram of Fig.\ \ref{Diagram_C2} and inserting the result into Eq.\ (\ref{density_density}) gives
\begin{eqnarray}
\label{C2_correlation}
&&\overline{\delta n(\textbf{r}, t) \delta n(\textbf{r}^{\prime}, t^{\prime})}
= \frac{2 \pi  \ell \hbar^2 A}{3 m^2 (2 \pi)^4}\times \nonumber \\
&&\int\limits_{-\infty}^{\infty} d z_1
\left[ \prod\limits_{j=1}^{2}  d k_j \; d \Omega_j \;
\left| \phi(k_j) \right|^2
P_{\epsilon_{k_j}}(z_1, \Omega_j) \right]
\nonumber \\
&&\times \partial_{z_1} P_{\epsilon_{-}}(z-z_1, \Omega_{-})
\partial_{z_1} P_{\epsilon_{+}}(z^{\prime}-z_1, \Omega_{+})
e^{-i(\Omega_{+} t + \Omega_{-} t^{\prime})},\nonumber\\
\end{eqnarray}
where $\epsilon_{\pm} = [\epsilon_{k_1} + \epsilon_{k_2} \pm
\hbar (\Omega_1 - \Omega_2)/2]/2$ and
$\Omega_{\pm} = \pm (\epsilon_{k_1} - \epsilon_{k_2}) + (\Omega_1 + \Omega_2)/2$.
This equation can be evaluated numerically and allows for analytical analysis in some special cases, as we now show.

\begin{figure}[t]
\includegraphics[width=6.3cm]{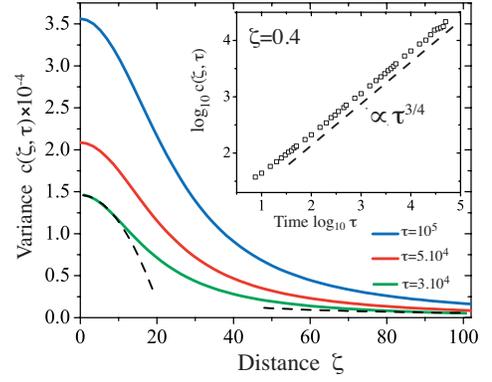}
\caption{\label{Curves_Fluc}
(color online). Position dependence of the correction $c$ to the variance of atomic density in an expanding BEC for three different times $\tau=t\mu/\hbar$. The dimensionless distance is $\zeta=z\sqrt{k_\mu\ell}/\ell$. The dashed lines show analytic results $B\tau^{3/4}[1 -C (\zeta/\tau^{1/4})^2]$ (small $\zeta$) and $D\tau/\zeta$ (large $\zeta$) for $\tau = 3 \times 10^4$. The numerical constants $B$, $C$ and $D$ given in the text. The inset shows the time dependence of $c$ for a given (small) $\zeta = 0.4$.}
\end{figure}

The simplest quantity that Eq.\ (\ref{C2_correlation}) allows us to study is the correction to the (normalized) variance of the atomic density fluctuations: 
\begin{eqnarray}
\label{Variance0}
\dfrac{\overline{\delta n^2(\textbf{r},t)}}{\bar{n}^2(\textbf{r},t)} =
1 + C_2(\textbf{r}, t; \textbf{r}, t),
\end{eqnarray}
where the unity on the right-hand side originates from the $C_1$ term and  
\begin{eqnarray}
\label{Variance}
C_2(\textbf{r}, t; \textbf{r}, t)=\dfrac{2}{k_\mu^2A \sqrt{k_\mu\ell}}c(\zeta,\tau)
\end{eqnarray}
with a combinatorial factor 2 added.
Here we introduced dimensionless variables $\zeta=z\sqrt{k_\mu\ell}/\ell$ and $\tau=t\mu/\hbar$ that feature natural spatial and temporal scales of the problem. The dependence of $C_2$ on position $\zeta$ and time $\tau$ appears to be given by a universal function $c(\zeta,\tau)$ that does not depend on any parameters of the problem. We plot this function in Fig.\ \ref{Curves_Fluc} for three fixed values of $\tau$. It is quadratic in $\zeta$ for small $\zeta\ll\tau^{1/4}$: $c(\zeta,\tau)\simeq B\tau^{3/4}[1 - C (\zeta/\tau^{1/4})^2]$ and decays as $D\tau/\zeta$ for $\zeta\gg\tau^{1/4}$. Here $B \simeq 6.3$, $C \simeq 0.3$ and $D \simeq 1.9$ are constants that had to be calculated numerically. $c(\zeta,\tau)$ grows with time as $\tau^{3/4}$ for $\zeta\ll\tau^{1/4}$ as we also show in the inset of Fig.\ \ref{Curves_Fluc}. This amplification of $c(\zeta, \tau)$ with time can make $C_2$ significant for large $\tau$, despite the small prefactor in front of $c$ in Eq.\ (\ref{Variance}). $C_2(\textbf{r}, t; \textbf{r}, t)$ becomes of order 1 for $t \sim (\ell/v_{\mu}) (k_{\mu} d)^{8/3} (k_{\mu} \ell)^{-1/3}$, which is still smaller than the localization time $t_{\mathrm{loc}}$ that limits the validity of our analysis.

\begin{figure}
\includegraphics[width=6.3cm]{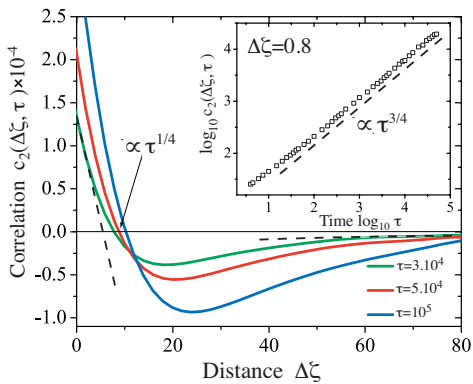}
\caption{\label{Curves_C2} 
(color online). Long-range correlation function $c_2$ versus dimensionless distance $\Delta \zeta=2\zeta$ (solid lines) for three different times $\tau=t\mu/\hbar$. The dashed lines show asymptotes $B\tau^{3/4}(1-E\Delta\zeta/\tau^{1/4})$ and $-F\tau/\Delta\zeta$ for $\tau = 3 \times 10^4$. The numerical constants $B$, $E$ and $F$ are given in the text. The inset shows the time dependence of the correlation for $\Delta\zeta = 0.8$.}  
\end{figure}

The long-range correlation of the fluctuations of atomic density is obtained from Eq.\ (\ref{C2_correlation}) with $\Delta z = \left| z - z^{\prime} \right| \gg \ell$. Note that $C_1$ correlation is negligible for such large spatial separations. In this work we restrict ourselves to equal-time correlations: $ t = t^{\prime}$. As an example, we consider correlations of density at two points symmetric with respect to the origin: $z = -z^{\prime} = \Delta z/2$, but results are qualitatively similar for any sufficiently distant $z$ and $z^{\prime}$ of opposite sign. We obtain
 \begin{equation}
C_2(\textbf{r}, t; \textbf{r}^{\prime}, t)=\dfrac{1}{k_\mu^2A \sqrt{k_\mu\ell}}c_2(\Delta\zeta,\tau).
\end{equation}
We show $c_2$ in Fig. \ref{Curves_C2} as a function of $\Delta \zeta$ for three different times.
For small distances $\Delta\zeta\ll\tau^{1/4}$, the correlation decays linearly with $\Delta\zeta$: $c_2(\Delta\zeta,\tau)\simeq B \tau^{3/4}(1 - E \Delta\zeta/\tau^{1/4})$, where $E \simeq 2$. At $\Delta\zeta \sim \tau^{1/4}$, it becomes \emph{negative} and reaches a minimum. For longer distances $\Delta\zeta \gg \tau^{1/4}$, $c_2$ remains negative whereas its magnitude decays only algebraically: $c_2 \simeq -F\tau/\Delta\zeta$ with $F \simeq 1.2$.
In addition to having long range is space, $C_2$ correlation grows in magnitude with time (see the inset of Fig.\ \ref{Curves_C2}), similarly to the variance of $\delta n$. This should facilitate its experimental observation. 

Negative correlations of atomic density in an expanding BEC could be anticipated from  the conservation of atom number $N$ which implies $\int d^3\textbf{r} \overline{\delta n(\textbf{r},t)\delta n(\textbf{r}^{\prime}, t)}=0$ and thus requires that the integrand must change sign. (Note that our Eq.\ (\ref{C2_correlation}) obeys this condition \emph{exactly}.) The important result of our work is to show that negative correlations occur at large distances between points $\textbf{r}$ and $\textbf{r}^{\prime}$ and that  they become increasingly important as the condensate expands.
Negative correlations of similar origin were predicted to exist in reflection of waves from a thick disordered slab \cite{rogozkin95}.
The analysis of long-range correlations introduces a new characteristic length scale $\zeta^* \sim \tau^{1/4}$ or $z^* \sim (D_{\mu} t \cdot \ell/k_{\mu})^{1/4}$ which is much smaller than the root mean square size of the condensate $\langle z^2 \rangle^{1/2}$. It determines the typical separation between two points, situated symmetrically with respect to the initial location of the condensate, at which density correlations change sign. 

In conclusion,  the macroscopic coherence of the condensate prevents the breakdown of the atomic speckle pattern $n(\textbf{r}, t)$ into small independent spots, imposing correlations between distant points. As a consequence, atomic speckles appear to have a much more complex and nontrivial spatial structure than just a random arrangement of small regions of high and low density put forward in Ref.\ \cite{Henseler}. Although in the present paper we consider an uncorrelated, white-noise potential, our results can be generalized to correlated potentials in a standard way \cite{Akkermans,kuhn07}.
We estimate that the long-range correlations that we study in this Letter should be directly observable under conditions of the experiment of Ref.\ \cite{Billy} for an \emph{isotropic} random potential with the correlation length equal to the longitudinal correlation length of Ref.\ \cite{Billy} and a slightly weaker transverse confinement. It would be extremely interesting to extend our calculation to the Anderson localization regime ($z > \xi_{\mu}$), where, by analogy with microwave experiments \cite{Chabanov}, one expects anomalously large density fluctuations.

We thank B.A. van Tiggelen and A. Minguzzi for discussions. S.E.S. acknowledges financial support of the French ANR (Project No. 06-BLAN-0096 CAROL) and the French Ministry of Education and Research.


\begin{thebibliography}{99}

\bibitem{experimental_papers} J.E. Lye  \emph{et al.}, Phys. Rev. Lett.
\textbf{95}, 070401 (2005); D. Clement \emph{et al.}, \emph{ibid}
\textbf{95}, 170409 (2005); C. Fort \emph{et al.}, \emph{ibid}
\textbf{95}, 170410 (2005). 

\bibitem{Sanchez-Palencia} L. Sanchez-Palencia \emph{et al.}, Phys. Rev. Lett.
\textbf{98}, 210401 (2007).

\bibitem{Shapiro} B. Shapiro, Phys. Rev. Lett.
\textbf{99}, 060602 (2007).

\bibitem{Sergey} S.E. Skipetrov, A. Minguzzi, B.A. van Tiggelen, and B. Shapiro, Phys. Rev. Lett.
\textbf{100}, 165301 (2008).

\bibitem{Billy} J. Billy \emph{et al.}, Nature
\textbf{453}, 891 (2008).

\bibitem{Roati} G. Roati \emph{et al.}, Nature
\textbf{453}, 895 (2008).

\bibitem{Henseler} P. Henseler and B. Shapiro, Phys. Rev. A
\textbf{77}, 033624 (2008).

\bibitem{goodman}
J.W. Goodman,
\emph{Speckle Phenomena in Optics} (Roberts \& Company Publishers, NY, 2007).

\bibitem{kane88}
C.L. Kane, R.A. Serota, and P.A. Lee,
Phys. Rev. B \textbf{37}, 6701 (1988).

\bibitem{feng91}
S. Feng and P.A. Lee,
Science \textbf{251}, 633 (1991).

\bibitem{Berkovitz}
R. Berkovits and S. Feng, Phys. Rep.
\textbf{238}, 895 (1994); M.J. Stephen and G. Cwilich, Phys. Rev. Lett.
\textbf{59}, 285 (1988); R. Pnini and B. Shapiro, Phys. Rev. B \textbf{39}, 6986 (1989).

\bibitem{Akkermans} E. Akkermans and G. Montambaux, \emph{Mesoscopic Physics of Electrons and Photons} (Cambridge University Press, Cambridge, 2007).

\bibitem{Pitaevskii} L. Pitaevskii and S. Stringari, \emph{Bose-Einstein Condensation} 
(Clarendon, Oxford, 2003).

\bibitem{Kagan} Yu. Kagan, E.L. Surkov, and G.V. Schlyapnikov, Phys. Rev. A
\textbf{54}, R1753 (1996);
Y. Castin and R. Dum, Phys. Rev. Lett.
\textbf{77}, 5315 (1996).

\bibitem{Gorkov} L.P. Gorkov, A. Larkin, and D.E. Khmelnitskii, JETP Lett.
\textbf{30}, 228 (1979).

\bibitem{Hikami} S. Hikami, Phys. Rev. B
\textbf{24}, 2671 (1981).

\bibitem{rogozkin95}
D.B. Rogozkin and M. Yu. Cherkasov,
Phys. Rev. B \textbf{51}, 12256 (1995).

\bibitem{kuhn07}
R.C. Kuhn \emph{et al.},
New. J. Phys. \textbf{9}, 161 (2007).

\bibitem{Chabanov} A.A. Chabanov, M. Stoytchev, and A.Z. Genack, Nature (London)
\textbf{404}, 850 (2000).

\end{thebibliography}
\end{document}